# High dynamic range open-loop current measurement based on diamond quantum magnetometer achieving ppm scale precision

*Qihui Liu, Hao Chen\*, Fei Xie, Yuqiang Hu, Jin Zhang, Nan Wang, Lihao Wang, Yichen Liu, Yang Wang, Zhichao Chen, Lingyun Li, Jiangong Cheng\* and Zhenyu Wu\**

Q. Liu, H. Chen, F. Xie, J. Zhang, N. Wang, L. Wang, Y. Liu, Y. Wang, J. Cheng, Z. Wu
State Key Laboratory of Transducer Technology, Shanghai Institute of Microsystem and Information Technology, Chinese Academy of Sciences, Shanghai, 200050, China
Email: haochen@mail.sim.ac.cn; jgcheng@mail.sim.ac.cn; zhenyu.wu@mail.sim.ac.cn

Q. Liu, H. Chen, F. Xie, N. Wang, Z. Chen, L. Li, J. Cheng, Z. Wu
University of Chinese Academy of Sciences, Beijing, 100049, China

Y. Hu, Z. Wu
School of Microelectronics, Shanghai University, Shanghai, 200444, China

Y. Hu, Z. Wu
Shanghai Industrial μTechnology Research Institute, Shanghai, 201800, China

Z. Chen, L. Li
Center for Excellence in Superconducting Electronics, Shanghai Institute of Microsystem and Information Technology, Chinese Academy of Sciences, Shanghai, 200050, China

Keywords: nitrogen-vacancy centers, quantum state manipulation, current sensor

**Abstract**

Negatively charged nitrogen-vacancy (NV) centers in diamond have been extensively studied as a promising high sensitivity solid-state magnetic field sensor at room temperature. However, their use for current sensing applications is limited due to the challenge of integration and miniaturization of the diamond NV sensor. Here, we demonstrate an integrated NV sensor fabricated with standard microfabrication process. The sensor device incorporated with a toroidal magnetic yoke enables a high-precision wide range direct current sensing with galvanic isolation. The performance of the diamond NV current sensor in an open loop configuration has been investigated. A current measuring range of 0 A ~ 1000 A with an uncertainty of 46 ppm are achieved. Taking advantage of dual spin resonance modulation, temperature drift is suppressed to 10 ppm/K. This configuration opens new possibilities as a robust and scalable platform for current quantum sensing technologies.

## 1. Introduction

Development of current sensors have gained momentum due to the strong need in many industrial applications, including electric vehicle,[1, 2] smart grid,[3, 4] and chemical industries.[5] In recent years, many high precision current sensor technologies based on different mechanisms have been studied. Hall-effect [6-8] and magneto resistance sensors[9, 10] - which are mass produced with semiconductor technologies - are the most widely used current sensors in the industry, thanks to the large dynamic range, low cost and low energy consumption. Nevertheless, their sensitivity are usually limited and are susceptible to temperature drift. Superconducting quantum interference devices (SQUID)[11, 12] and optically pumped magnetometers[13, 14] based on quantum effects provide an extension for current sensors due to their sub-femto-Tesla sensitivity.[15] However, the dynamic range is limited to ~mT and the working condition — such as the need of a cryostat or a thermal chamber - hinders their use as current sensors. The nitrogen-vacancy (NV) center in diamond is a solid-state quantum sensing platform that attracted extensive attention in recent years.[16, 17] The defect's spin energy levels are sensitive to magnetic fields, electric fields, strain and temperature, allowing NV to operate as a multimodal sensor. Quantum sensing methods based on NV centers are characterized by high magnetic sensitivity, large dynamic range, and high spatial resolution under ambient conditions,[18, 19] and have been used to measure temperature,[20, 21] biomolecules[22, 23], and pressure[24], as well. Recent advances have demonstrated the feasibility of NV sensors for current measurement and automotive batteries,[25, 26] but it requires compact and scalable architectures for practical use to replace bulky and complex laboratory instruments.

In this work, we demonstrate an open-loop current sensing mechanism incorporated with a microfabricated diamond NV sensor and a high permeability magnetic yoke. Section 2 describes the principle and experimental setup of the diamond NV sensor. In analogue to conventional open-loop Hall current transducer configuration, the diamond NV sensor, which is a magnetic field sensor for the flux change induced by the primary current is placed in the gap of a magnetic yoke. Pump laser and Microwave (MW) is delivered through standard fiber optics and coaxial cable allowing spin polarization and readout. The experimental measurement of the sensor is described in section 3. We investigated the performance of the diamond NV current sensor, including linearity, backhaul errors, repeatability errors and uncertainty. The minimum detection limit is below the noise of the current source in use. To suppress the temperature drift, dual resonance modulation scheme is employed.

## 2. Principle and Method

### 2.1 Quantum states and measurement principles of NV sensor

NV center is a color center composed of a negatively charged nitrogen atom and a vacancy spot in diamond. The sensing mechanism of NV centers relies on the interaction between magnetic fields and electron spins. Considering the nitrogen's nuclear spin (I = 1 for $^{14}$N) of the NV center, ground state Hamiltonian of the NV center in the applied magnetic field B can be expressed as

$$\mathcal{H}_S \approx D_{gs}S_z^2 + g_s\mu_B\vec{B}\cdot\vec{S} + A_\parallel S_z I_z + A_\perp(S_x I_x + S_y I_y) \qquad (1)$$

where $D_{gs} \approx 2.87$ GHz is the NV zero-field splitting, $\mu_B$ is the Bohr magneton, $g_s \approx 2.0028$ is the approximately isotropic g-factor, and $\vec{S} = (S_x, S_y, S_z)$ is the dimensionless electronic spin-1 operator.[27] The first two terms encompass the NV− electron spin interaction with external magnetic field and temperature variation. The last two terms are derived from the additional hyperfine energy splitting of the $^{14}$N nucleus. $A_\parallel$ = −2.158 MHz and $A_\perp$ = −2.70 MHz are the axial and transverse magnetic hyperfine coupling coefficients. **Figure 1**a shows the electronic ground state ($^3A_2$) and the excited states ($^3E$) of the NV center spin S =1, which are both spin triplet states. The intensity of photoluminescence (PL) follows its spin-level-dependent path: spin states with $m_s$ =±1 cause a decrease in PL due to a non-radiative relaxation path from the singlet to the ground state ($^1A_1$). In presence of a magnetic field B, a shift of the resonant frequency ($f_\pm$) of the spin transition of the NV center due to Zeeman splitting is:[28]

$$f_\pm \approx D_{gs} + \beta_T \delta_T \pm \gamma_{NV} B_{NV} \qquad (2)$$

Where $\beta_T \approx 74.2$ kHz/K near room temperature,[29] $\delta_T$ is the temperature offset from 293 K during the measurement, $\gamma_{NV} = g_s\mu_B/h \approx 28$ Hz/nT is the gyromagnetic ratio, and $h$ is the Planck constant, $B_{NV}$ is the projection of the magnetic field B along the NV symmetry axis. By calculating the difference in resonant frequencies, $\Delta f$ is proportional to the magnetic field, and compensates the effects of temperature drift:

$$\Delta f = f_+ - f_- \approx 2\gamma_{NV} B_{NV} \qquad (3)$$

The diamond current sensor consists of diamond chip, microwave (MW) printed circuit board (PCB), fiber coupling module (FCM), and magnetic yoke, as shown in Figure 1(b). A diamond plate with a natural abundance of $^{13}$C isotope (3×3×0.5 mm$^3$,

[$^{14}$N] < 13 ppm, [100] orientated) is encapsulated in the silicon chip with an etched aperture for laser excitation through FCM. The MW PCB soldering with the diamond chip to offers an access for MW field. The diamond sensor device is mounted in the gap of a magnetic yoke, where the magnetic field generated by the primary current in the cable concentrates. The magnetic field is sensed by the NV centers in the diamond. At resonance, the microwave field drive the electrons in the ground state population transferred from $|m_s = 0\rangle|$ to $|m_s = \pm1\rangle$ sublevels. The right side of Figure 1a shows the interaction of the NV center with the I = 1 nuclear spin leads to a splitting of the given energies into triplets.[30] Three applied frequencies spaced at 2.158 MHz resonantly address fine energy levels, which exhibit five characteristic zero-crossings on the optically detected magnetic-resonance (ODMR) spectrum for accuracy improvement.[31] The ODMR spectrum of the NV ensemble recorded under the bias magnetic field is presented in Figure 1(c). Four pairs of resonant transition frequencies can be observed corresponding to the NV axis in the four possible crystallographic directions, respectively. The resonant frequency of NV$_4$ at which the magnetic field is maximally projected on the NV axis is tracked to ensure maximum sensitivity.

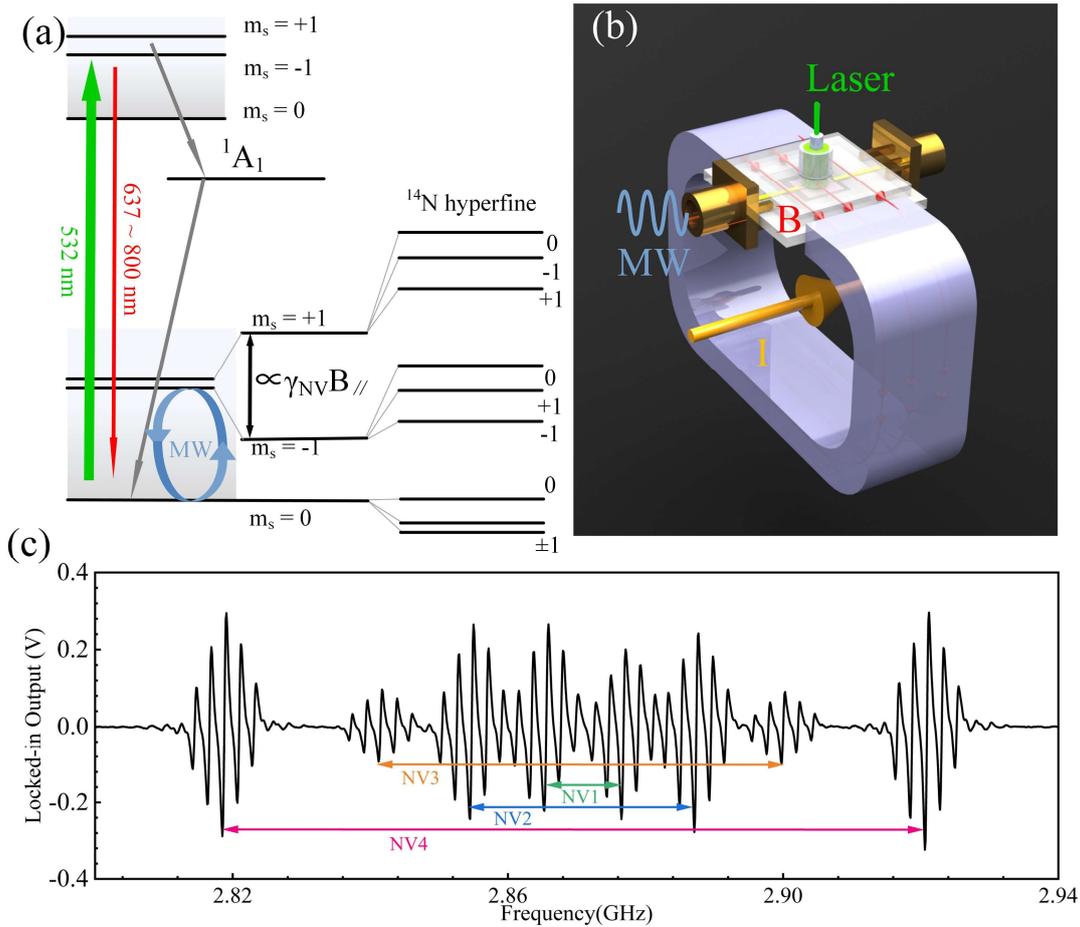

**Figure 1** a) Energy-level diagram for the NV center in diamond, with two spin triplet states (fundamental and excited) and one singlet metastable state. The right side shows the hyperfine energy levels coupled with the spin of the $^{14}$N nucleus. b) Schematic architecture of the diamond NV current sensor, driven by modulated MW. c) Initial ODMR spectrum observed with current I=0 with a bias magnetic field. There are five characteristic zero-crossings at each resonance frequency, which is the simultaneous manipulation of all three $^{14}$N hyperfine resonances.

## 2.2 Experimental setup and device fabrication

**Figure 2**a depicts the diagram of the experimental setup. A current carrying conductor creates a magnetic field. This field is concentrated by a magnetic yoke. The yoke has a gap cut through it and a diamond NV sensor is inserted to sense the magnetic flux density ***B*** in the gap. Two separate microwave sources generate two carrier signals, which correspond to the pair of ODMR features of NV$_4$ in Figure 1(c). Both of MW tones are square-wave frequency modulated at frequencies of $f_1$ = 8.5 kHz, $f_2$ = 11 kHz, with the same frequency deviation $\delta v$ = 550 kHz. An RF source produce ±2.158 MHz sidebands to match the $^{14}$N hyperfine resonances. Synthesized signal is delivered to the diamond NV ensemble to drive spin transition. A 532nm laser is coupled to the fiber circulator and focus on the diamond through FCM. The fluorescence is reflected into the circulator and directed through a 633-nm long-pass filter, after which 16 μW of PL is collected by photodiode (PD). In addition, the beam splitter picks up a portion of laser power as a reference for common mode noise rejection (CMR). The result signal is demodulated by the lock-in amplifier. Microwave source is tuned by a closed-loop program to track the variation of the current. Figure 2b shows the setup of the current sensor, therein the integrated diamond sensor is fixed by a 3D printed housing and the current flow in the cable penetrating through the magnetic yoke is measured. For microwave-based spin manipulation, a coplanar waveguide (CPW) is implemented on PCB. Figure 2c shows more details on the rear side of the integrated diamond sensor.

The diamond chip is fabricated by stacking three layers of silicon wafers. Figure 2d - 2f show the schematic side views of stacking in sequence, as shown in Figure 2h-2j. The process consists of three thermal compression bonding processes and the diamond is fixed in the chip and protected. Figure 2(k) is the backside of Figure 2(j) with an opening which provides an access to the fiber module. A single diamond chip (Figure 2(g)) can be obtained by dicing the wafer after the process is completed. Detailed fabrication process of the sensor has been reported elsewhere.[32]

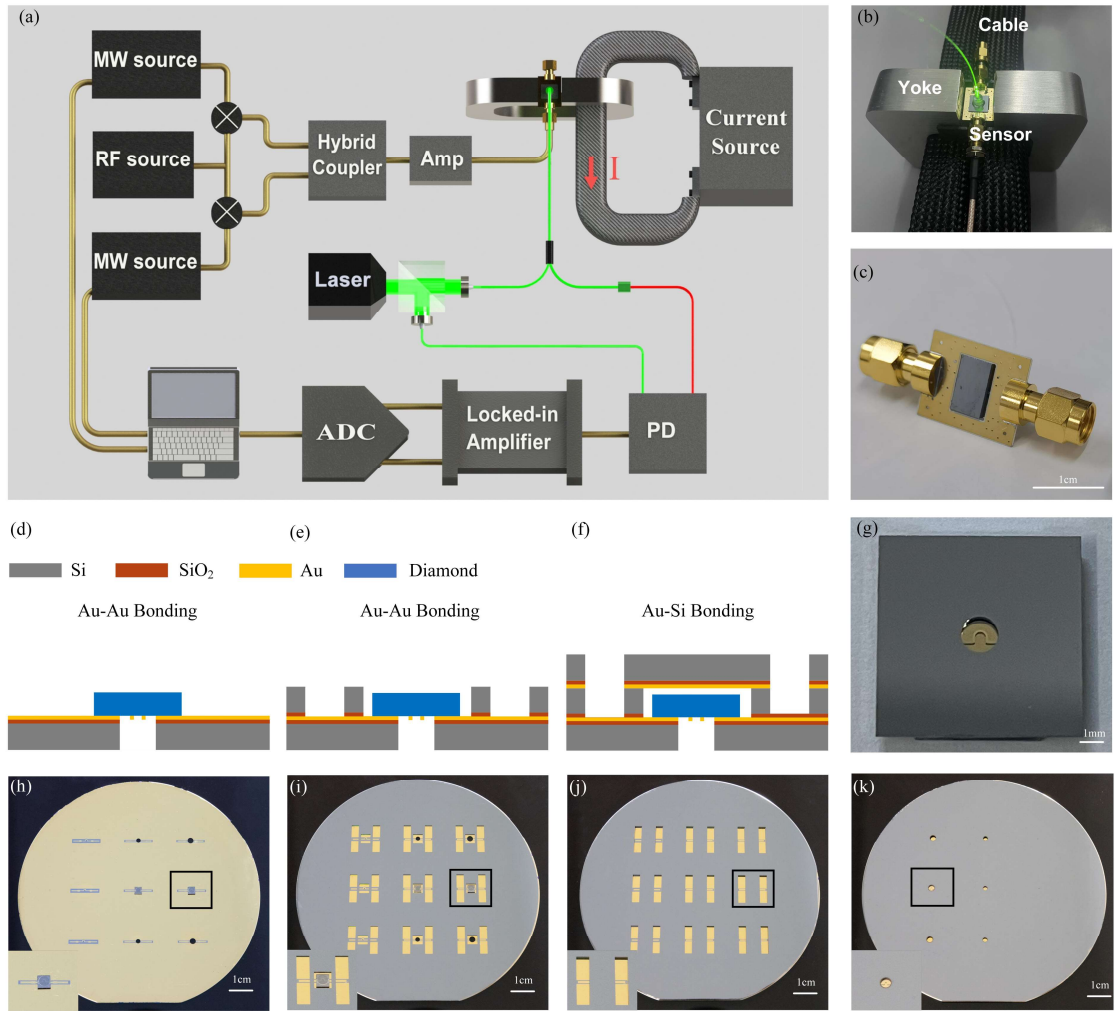

**Figure2** a) Schematic of the setup. MWs are mixed with a radiofrequency (RF) signal at 2.158 MHz to produce modulated carriers, which are amplified, combined, and delivered to drive the sensor. The beam passes through a beam splitter, where a fraction of laser is coupled into the fiber circulator and the rest of the beam is collected by a PD. The diamond PL is reflected back to the fiber circulator and imaged onto the PD after long-pass filtering at 633nm. The differentially amplified signal of the PD is demodulated by lock-in amplifier, and then processed in the close-loop system. b) Photograph of the configuration of the current sensor. c). The rear side view of the diamond sensor. d)-f) Sequentially illustrate the three-steps of bonding process for diamond sensor micromachining (g). Photograph of a single diamond chip, the omega pattern produces a uniform microwave field at the center. (h)-(j) Corresponding photograph of (d)-(f). The inset is the chip used for fabricating the current sensor after dicing. (k) The rear side view of (j).

## 3. Results and Discussion

### 3.1 Sensitivity and linearity

Diamond NV current sensor employs a high permeability magnetic yoke to detect magnetic flux density produced by a current flow through a conductor. Magnetic yoke can amplify magnetic fields and are therefore valuable in improving accuracy, resolution, and linearity. The soft ferromagnetic yoke we used in this study is Nanocrystalline alloy yoke (1K107, relative permeability of 50000-80000, 59×36×9 mm$^3$) and steel yoke (Q235A, relative permeability of 7000-10000, 108×88×15 mm$^3$). The same diamond NV sensor is used for both yokes for the subsequent study. **Figure 3**a demonstrates the sensitivity and linearity of diamond NV sensor with Nanocrystalline alloy yoke (described as sensor 1 in the following text). The sensitivity is described by the slope of the characteristic magnetic field output. A sensitivity of 30491 nT/A is obtained at a measurement range of 0 to 300 A. After the current exceeds 300 A, the magnetic flux saturation of the magnetic yoke leads to an unambiguous nonlinearity. As a result, the linear range of the measurement is restricted below 300 A for sensor 1. Error bars in black demonstrate the standard deviation for magnetic field measurements of 1 s data with 10 kHz sampling rate. The most significant error occurs at 100 A owing to the shifting gears (100 A to 1000 A) of the output range for the current source. The curve in Figure 3b is obtained by diamond NV sensor with steel yoke installed (described as sensor 2 in the following text). A lower sensitivity of 3326 nT/A at a measurement range of 1000 A (the upper limit of load for the cable in use) is obtained. For the open loop configuration, the measurement range is restrained by the magnetic saturation for the magnetic yoke and the dynamic range of the diamond NV sensor. As the diamond NV sensors based on Zeman effects has a broad range up to a few Tesla and above, the dynamic range is determined by the saturation characteristics of magnetic core material.[33, 34] As a result, the measurement is confined within the linear region of the hysteresis loop of the material in use. Sensitivity can be further improved by optimizing the size of the yoke gap[6]. Figure 3c,d show the linearity characterization of the sensor and the nonlinear error of the sensor is described as

$$\delta_L = \frac{|\Delta L_{max}|}{Y} \qquad (4)$$

where $\Delta L_{max}$ is the largest difference between the data points and fitting curve, Y is the dynamic range of the current sensor. As shown in Figure 3a, the calculated nonlinear errors of sensor 1 and 2 are 0.11% and 0.05%. The nonlinear error attributes to the nonlinear transfer curve of the magnetic flux in the magnetic yoke with increasing current due to saturation and hysteresis, which is the general drawback for open loop configuration.

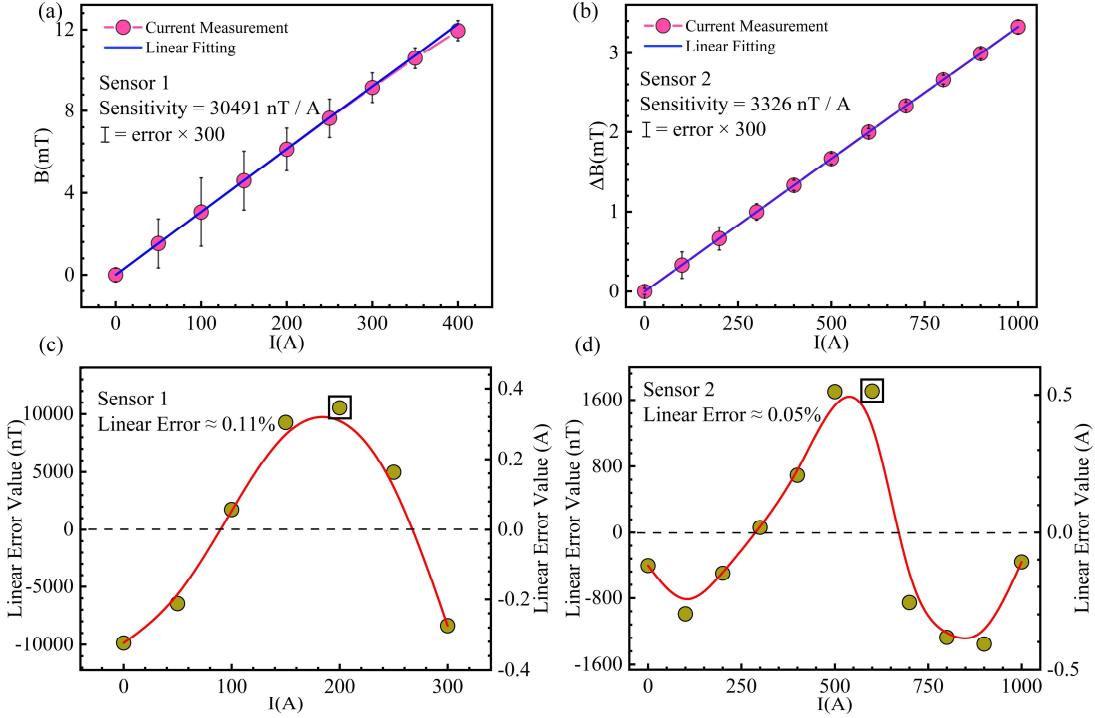

**Figure 3** Linear fitting curves (solid blue line) of dc current measurement of a) sensor 1, b) sensor 2, the pink line connects the measuring points. The error bars represent standard deviation of the output signal and are exaggerated 300 times for clarity. The variation of the magnetic flux density induced by the current in the cable is determined by the characteristic zero-crossing frequency of the $NV_4$ axis. c,d) The extracted nonlinear error for sensor 1 and sensor2. The black boxes represent the points with the maximum offset.

### 3.2 Hysteresis and noise floor

Despite working in the linear range of the hysteresis loop, backhaul error is affected by the remanent magnetization of yoke material. For sensors with magnetic yokes, the backhaul error and repeatability error must be kept within a reasonable range (3%) to ensure measurement accuracy.[35] The backhaul error and repeatability error are derived from:

$$\delta_H = \frac{|\Delta H_{max}|}{Y} \quad (5)$$

$$\delta_R = \frac{|\Delta R_{max}|}{Y} \quad (6)$$

where $\Delta H_{max}$ is the maximum deviation between the curves of forward travel and reverse travel. $\Delta R_{max}$ is the largest difference among multiple codirectional measurement curves. In **Figure 4**a,b, the difference between the fronthaul path and the return path of sensor 1 and 2 is measured. Due to the hysteresis effect of the magnetic yoke, the return error reaches the maximum in the middle travel of the measurement is 0.06% (0.23%). In Figure 4c,d, the fronthaul path and the back path of sensor 1 and 2 were measured three times and the maximum deviation of the same direction path was calculated to scale the error interval of the repeatability measurement. The positive and negative signs were used to distinguish the fronthaul paths and backhaul paths and repeatability error of 0.027% and 0.025% are calculated. Bias instability refers to the random variation in the calculated deviation within a specified finite sampling time and average time interval is given by the bottom of the Allen deviation curve in Figure 4e,f, which represents the minimum detectable current. The Allen deviation provides the information of the correlation for successive measurements independent of overall (long-term) drift [18], as shown in Figure 4e,f. In the high frequency range, the Allan deviation approximately follows $t^{-1/2}$ scaling up to 0.5 s, and the noise coefficient can be represented by the intersection of the extension of the Allan deviation curve -1/2 slope line with the $t$ = 1s line. Bias instability refers to the random variation within a specified finite sampling time and average time interval is marked in Figure 4e,f, which represents the minimum detectable current. The noise performance of the devices were measured in Figure 4g,h. The voltage fluctuation of the sensor 1 and 2 for 1 s is measured in real-time at the sampling rate of 10 kHz and converted into current noise density spectral by Fourier transform and the noise density spectrum is averaged by repeating 100 times. The solid red represents the noise spectral densities measured at magnetically sensitive. The noise floor of sensor 1 and 2 is 0.24 mA·Hz$^{-1/2}$ and 2.1 mA·Hz$^{-1/2}$ at 1 Hz, which corresponds to the intersection of the Allen deviation curve with $t$ = 1 s in Figure 4e,f. In the magnetically sensitive low-frequency region, 1/$f$ noise is the main influencing factor, which is most possibly the result of the interference of the unshielded test environment. The noise of 50 Hz and its odd higher order terms attributes to the magnetic field generated by the power transformer.[36] The blue line represents the measured noise spectral density at magnetically insensitive frequencies, and the solid yellow line represents the electronic noise floor of the electronic device.

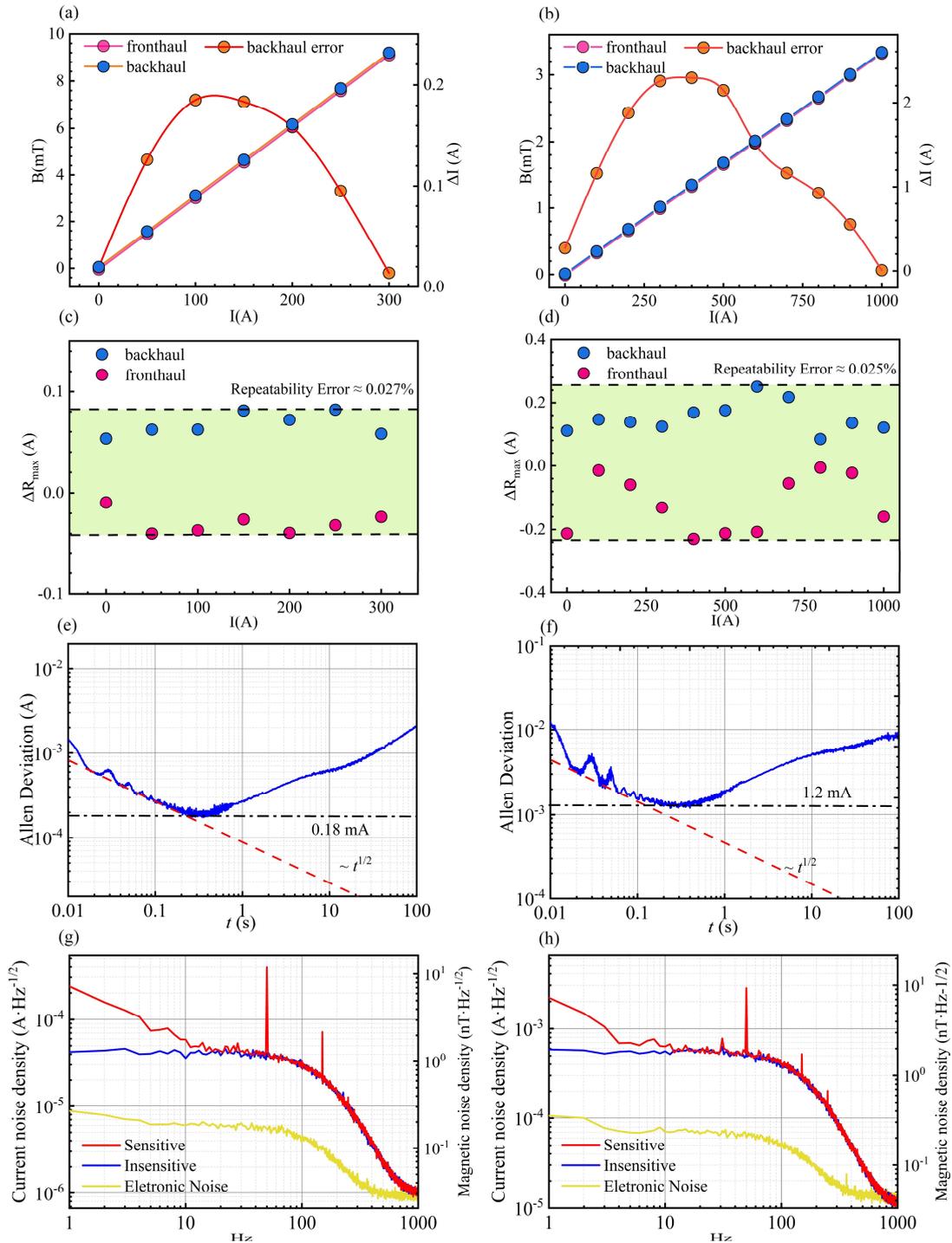

**Figure 4** Performance of current sensors 1 and sensor 2. a,b) Backhaul errors of sensor1 and 2, the orange dots is the difference between the fronthaul and backhaul paths. c,d) Repeatability error of sensor 1 and 2. The pink point is the maximum deviation of all fronthaul paths, and the blue point is the maximum deviation of all back paths. The positive and negative signs distinguish the fronthaul and backhaul paths. e,f) Allen deviation of sensor 1 and 2. The slope of the red dotted line indicates the desired scaling behavior to approach a central limit. g,h) Current noise spectra with magnetic sensitivity (red) and insensitivity (blue), and electronic noise (yellow) of sensor 1 and

2.

### 3.3 Performance of sensor

To determine the uncertainty of the current measurement, the fluctuation of the current over time is observed at different current intensity. In **Figure 5**, 2 s of data are recorded by NV sensors at a sampling rate of 10 kHz. Commercial close loop (LF1005S) and open loop (HAT1000S) Hall current sensors (LEM) are implemented to do the measurement simultaneously as a control group. It turns out that the fluctuation is not only originated from the sensor but also from the instability of the current source equipment. We analyzed the noise level by recording the signal as a time trace to monitor the variation when the source is switched off. From the measured current variation, shown in Figure 5a, it can be clearly identified when the source power is turn off at t=1.0 s. The deviation of the signal is reduced from 170 mA to 7 mA and 46 mA for sensor 1 and 2, respectively. This is direct evidence that the detection limit of the diamond NV current sensor 1 and 2 is below the noise level of the source (~170mA). In the following measurement, the noise of the source is dominant. Therefore, we take the maximum deviation captured by the sensor as the sensing uncertainty when the source power off. An uncertainty of 23 ppm and 46 ppm for sensor 1 and 2 are obtained. For Hall current sensors, the noise of the current source cannot be resolved. In such open-loop configuration, current sensors are affected by the external magnetic field's variation and nearby currents due to the magnetic leakage associated with the airgap in the magnetic circuit. Figure 5b,c show the current measurement of 100A and maximum capacity current, respectively. Figure 5d shows the real-time tracking a typical current change from 200A to 201A. The rise time is ~30ms, and the tracking rate of the current can be improved by reducing the switching time of the microwave (MW) and the demodulation time of the lock-in amplifier. This is discussed in detail published elsewhere.[37] The noise of the current source at 201A was evaluated by fitted Gaussian distribution (black curve), and sensor 1 and 2 show a consistent standard deviation of $\sigma \approx 0.12A$, which is a direct evidence that the current source contributes the dominant noise.

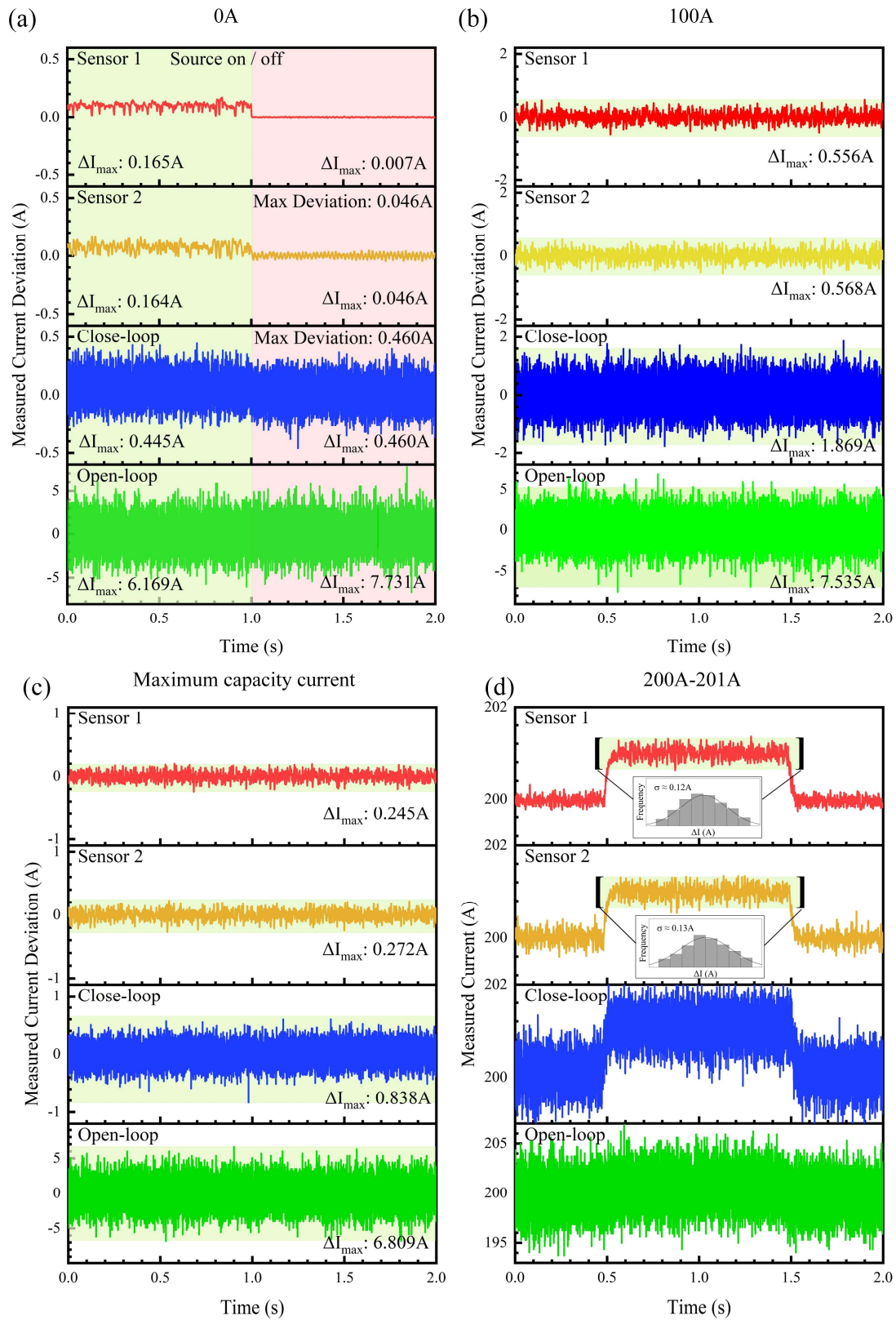

**Figure 5.** Performances of the NV sensor 1 and 2, compared with commercial close-loop and open-loop current sensors in different condition. a) 0 A current (source turned on and off); b) 100A current; c) maximum capacity current of 300 A and 1000A for sensor 1 and 2, respectively; d) Time traces of current change from 200A to 201A.

Current source on and off are distinguished by cyan and pink areas.

### 3.4 Temperature drift suppression

Temperature drift is a universal challenge for most current transducer. Aside from the embedded temperature compensation module for demodulation electronics, more effort has been put in to exclude the influence of temperature by investigating the sensing mechanism of sensor per se. Dual-frequency-modulation scheme is implemented in this work to mitigate temperature drift. According to **Equation 3**, a pair of resonance frequencies $f^{\pm}$ on the NV axis are simultaneously tracked and therefore the frequency difference can be readily calculated. Due to the temperature induced frequency shifts are codirectional for both $f^+$ and $f^-$, the thermal drift is reasonably cancelled out by extracting the term independent of temperature. We demonstrate the uncalibrated and calibrated current fluctuation by tracking one single resonance frequency and dual resonance frequencies, respectively. As shown in **Figure 6**, a slow drift of the temperature ~0.725°C increase is observed in a time range of 100 s which is possibly resulted from the laser radiation on the surface of the diamond. Real-time current intensity output with/without temperature calibration are plotted in Figure 6a,b for nanodiamond alloy core and steel core, respectively. The measured current deviation induced by temperature variation is specified as the temperature drift coefficient, and the temperature coefficients with/without temperature calibration for 4 current sensors tested in this work are listed in **Table1**.

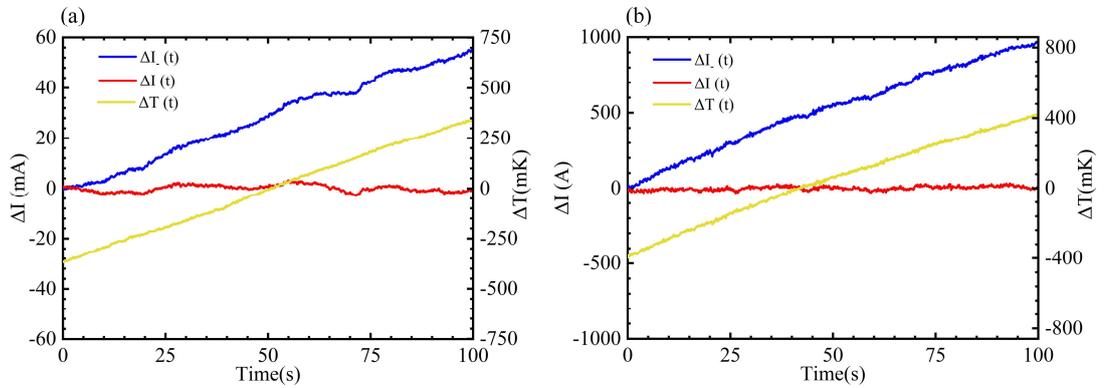

**Figure 6** a,b) Sensor 1 and 2, the temperature drift of localized single NV resonance frequency (blue solid) and suppression of localized double NV resonance frequencies (red solid) in 100s. The solid yellow line shows the temperature rise caused by pump laser irradiation.

Some typical parameters of sensor 1, sensor 2, commercial close loop and open loop Hall current sensors are summarized in Table 1. This indicates that the NV current sensors based on the quantum state manipulation demonstrate a superior performance and provides a high-precision current sensing solution.

**Table 1. Performance of different sensors.**

| Sensor | $\Delta I_{max}$ [a)][mA] | Range[A] | Linear Error | Bias Instability[mA] | Temperature coefficient[ppm/K] |
|---|---|---|---|---|---|
| Sensor 1 | 6.8 | 300 | 0.11% | 0.2 | 10(256[b)]) |
| Sensor 2 | 46.2 | 1000 | 0.05% | 1.2 | 22(1193[b)]) |
| LF1005S | 460 | 1000 | 0.10% | 1.74 | 83 |
| HAT1000S | 7731 | 1000 | 0.92% | 7.20 | 840 |

[a)] Measurements when the current source is turned off

[b)] The temperature coefficient without calibration

**Conclusion**

In this work, an integrated NV sensor based on a standard microfabrication process is proposed for current measurement. Our microfabrication-based current sensing scheme provides a scalable and flexible solution. The fabricated NV chip is assembled with PCB and optical fiber to form a compact solid-state device, which exhibit excellent precision and good temperature drift suppression. With the magnetic yoke properly selected, our prototype achieves a measurement range up to 1000A and uncertainty of 46 ppm. The temperature drift coefficient is reduced to 10 ppm by applying dual frequency modulation scheme. Recent research has verified the sensitivity of the diamond NV can be further improved to sub-pT $\cdot$Hz$^{-1/2}$ by material engineering[38] and sensor architecture design,[39] which would push the diamond NV based current sensor for further precision improvement. The open loop configuration presented in this work is prone to suffer from nonlinear errors due to flux saturation as discussed in the text. In future design, a secondary winding with close loop current feedback control can be added to make the equivalent magnetic flux in the yoke at a neutral level, so that the sensor measurement range and linearity can be further optimized. This work paves the way for extending the application of diamond NV current sensing from laboratory setups into practical applications such as on-chip current sensing, grid and control systems, and electric vehicles in the future.

**Acknowledgments**

The authors thank Professor Xiaohong Ge from the Shanghai Institute of Microsystem and Information Technology, Chinese Academy of Sciences (CAS), for sample preparation and useful discussion. The authors acknowledge the support from the grant from the CAS Strategic Pilot Project (No. XDC07030200), the R&D Program of Scientific Instruments and Equipment, Chinese Academy of Sciences (No. YJKYYQ20190026), and the National Key R&D Program of China (No. 2021YFB3202500).

DATA AVAILABILITY

The data that support the findings of this study are available from the corresponding authors upon reasonable request.